\documentclass[superscriptaddress,12pt,nofootinbib]{revtex4-2} 

\usepackage{mathtools}

\usepackage{booktabs,makecell,array}
\newcolumntype{P}[1]{>{\raggedright\arraybackslash}p{#1}}

\usepackage{amsmath}  
\usepackage{amsfonts} 
\usepackage{graphicx} 
\usepackage{esvect}
\usepackage{mathtools}

\usepackage{xcolor}
\usepackage{appendix}
\usepackage{siunitx} 
\usepackage{mathrsfs}

\usepackage{placeins}

\usepackage[caption=false]{subfig} 

\usepackage{pbox}

\usepackage{mathtools}
\makeatletter
\newcommand{\vast}{\bBigg@{3}}
\newcommand{\Vast}{\bBigg@{3.5}}
\makeatother

\begin{document}


\title{Self-Image Multiplicity in a Concave Cylindrical Mirror}


\author{Thach A. Nguyen}
\thanks{These two authors contributed equally.}
\affiliation{University of South Florida, Tampa, FL 33620, USA.}

\author{Kaitlyn S. Yasumura}
\thanks{These two authors contributed equally.}
\affiliation{Scripps College, Claremont, CA 91711, USA}

\author{Duy V. Tran}
\affiliation{Department of Mechanical Engineering, VNUHCM University of Technology, \\ Binh Thanh, Ho Chi Minh 700000, Vietnam.}

\author{Trung V. Phan}
\email{{tphan@natsci.claremont.edu}}
\affiliation{Department of Natural Sciences, Scripps and Pitzer Colleges, \\ Claremont Colleges Consortium, Claremont, CA 91711, USA}

\begin{abstract}
Concave mirrors are fundamental optical elements, yet some easily observed behaviors, such as the formation of multiple reflected images, are rarely addressed in standard textbooks. Here we investigate self-imaging -- where the observer is also the observed object -- using a concave cylindrical mirror. We predict the number of self-images visible from different observation points and classify the observation space into regions by image count. We then test these predictions with an inexpensive stainless-steel concave cylindrical mirror commonly found in teaching labs. This activity links geometrical optics principles to direct observation and provides a ready-to-use classroom demonstration and student exercise.
\end{abstract}

\date{\today}


\maketitle 

\section{Introduction}

\begin{figure*}[htbp]
\centering
\includegraphics[width=\textwidth]{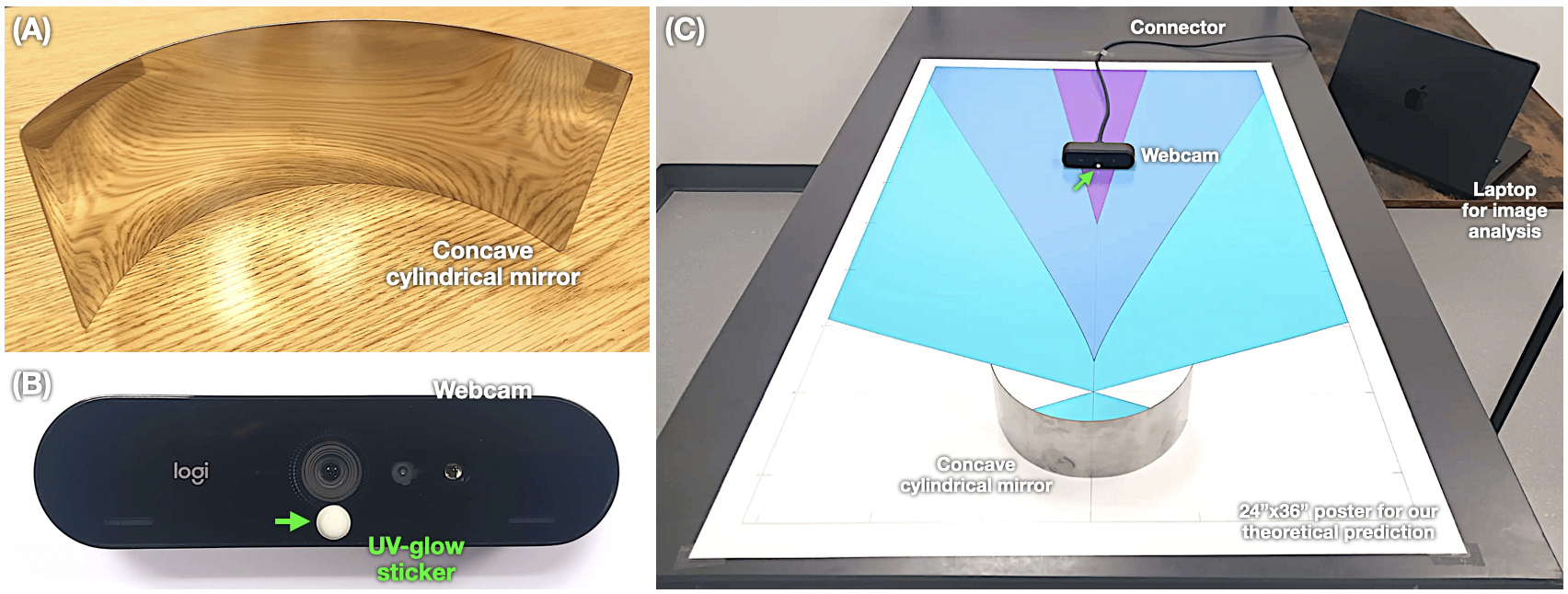}
\caption{\textbf{Experimental setup for observing self-images in a concave mirror.} Green arrows mark the light source. {\color{black}\textbf{(A)} Concave cylindrical mirror. \textbf{(B)} Webcam observer with a small marker (a UV-glow sticker) placed near the lens to co-locate the source and observer. \textbf{(C)} A 24''$\times$36'' poster marks the predicted regions for different self-image multiplicities; placing the webcam in these regions yields the corresponding number of glowing dots (see Fig. \ref{fig04}).}}
\label{fig01}
\end{figure*}

Many years ago, during an introductory physics lab on optics\footnote{PHY 104: Introductory Electromagnetism, Princeton University (Spring 2017)}, a group of students encountered an interesting dilemma. While exploring a set of mirrors intended to reinforce the lecture, they noticed something unexpected: with a concave mirror, their own reflections sometimes appeared as \textit{multiple images}, depending on where the mirror was placed. This curious phenomenon, unfortunately, had not been covered in their lecture and is still largely absent from standard textbooks \cite{halliday2010physics, jewett2008physics, young1996university}. Although a qualitative explanation satisfied the students at the time, we later realized that this simple observation did not have a \textit{detailed quantitative analysis} in the existing literature. The underlying physics is relatively simple, yet a careful investigation -- grounded in elementary geometry and basic image formation principles -- offers clear pedagogical and conceptual value. In this article, we study the phenomenon of \textit{self-image} formation -- in which the observer and the observed object are the same -- first theoretically using a \textit{two-dimensional} concave mirror in Section \ref{sec:theo}, and then experimentally using a \textit{three-dimensional} cylindrical concave mirror (see Fig. \ref{fig01}) in Section \ref{sec:exp}. The geometric measurements of the mirror used in our experiment are provided in Appendix \ref{app:measurement}. The experimental setup is straightforward and can be easily implemented in introductory laboratories.

\section{Two-Dimensional Theoretical Analysis \label{sec:theo}}

\subsection{On the Formation of Self-Image}

\begin{figure*}[htbp]
\centering
\includegraphics[width=\textwidth]{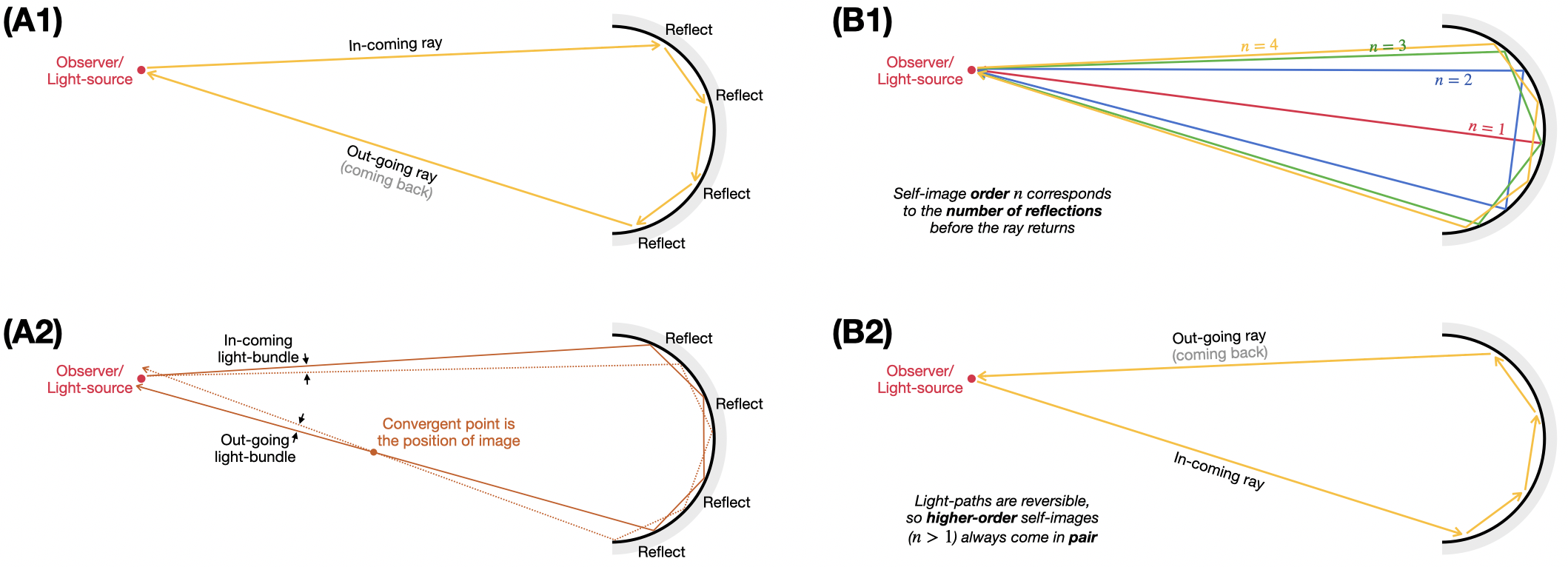}
\caption{{\color{black}\textbf{Light-ray paths for self-image formation.} The observer/light source is the red point. \textbf{(A1)} A ray leaves the observer, undergoes multiple reflections in the concave mirror, and returns to the observer. \textbf{(A2)} The corresponding image location is found from a narrow ray bundle by locating where the outgoing bundle converges. \textbf{(B1)} Rays may undergo many reflections; the number of reflections defines the self-image \textit{order} $n=1,2,3,...$. \textbf{(B2)} Reverse path of the higher-order ray in (A1).}}
\label{fig02}
\end{figure*}

Let us recall that a point-like observer sees a given self-image only if \textit{both} of the following criteria are satisfied for that image:
\begin{itemize}
    \item \textit{Comeback condition:} there exists a ray that leaves the observer, reflects on the mirror (possibly multiple times), and returns to the same point (see Fig. \ref{fig02}A1).
    \item \textit{Visibility condition:} for that same ray, the infinitesimal \textit{outgoing} bundle that accompanies it after reflection converges to a point located in front of the observer; this convergence point corresponds to the apparent position of the self-image (see Fig. \ref{fig02}A2).
\end{itemize}
In other words, each self-image corresponds to a \textit{comeback} ray that also satisfies the \textit{visibility condition}; if either condition fails, that self-image does not appear. Let us define the \textit{order} $n$ of a self-image as the number of reflections the corresponding light-ray undergoes before returning to the observer, then $n=1,2,3,...$ (see Fig. \ref{fig02}B1). Higher-order self-images ($n>1$) appear in pairs because of path reversibility (see Fig. \ref{fig02}B2).

The \textit{comeback condition} can be analyzed in a \textit{two-dimensional} concave mirror model, which makes the image-counting geometry apparent (as illustrated in Fig. \ref{fig02}). The \textit{visibility condition}, however, must be treated in \textit{three dimensions} because a cylindrical mirror is \textit{astigmatic}: horizontal (\textit{transverse}) and vertical (\textit{sagittal}) rays focus at different locations, so the small outgoing bundle may not form a single sharp focus in front of the eye \cite{hecht2012optics}. To keep the explanation simple, we defer visibility, astigmatism, and their consequences for image sharpness to Appendix \ref{app:astigmatism}. There we show that, because the \textit{sagittal} focusing point is always visible, the \textit{comeback condition} alone suffices for image counting.

\subsection{Comeback Condition and Spatial Partition with Self-Image Count \label{sec:comeback}}

In this Section, for each observer position, we identify every comeback ray, since each distinct comeback ray corresponds to one self-image. Using these counts, we partition the space into regions distinguished by the number of visible images.

\begin{figure*}[htbp]
\centering
\includegraphics[width=\textwidth]{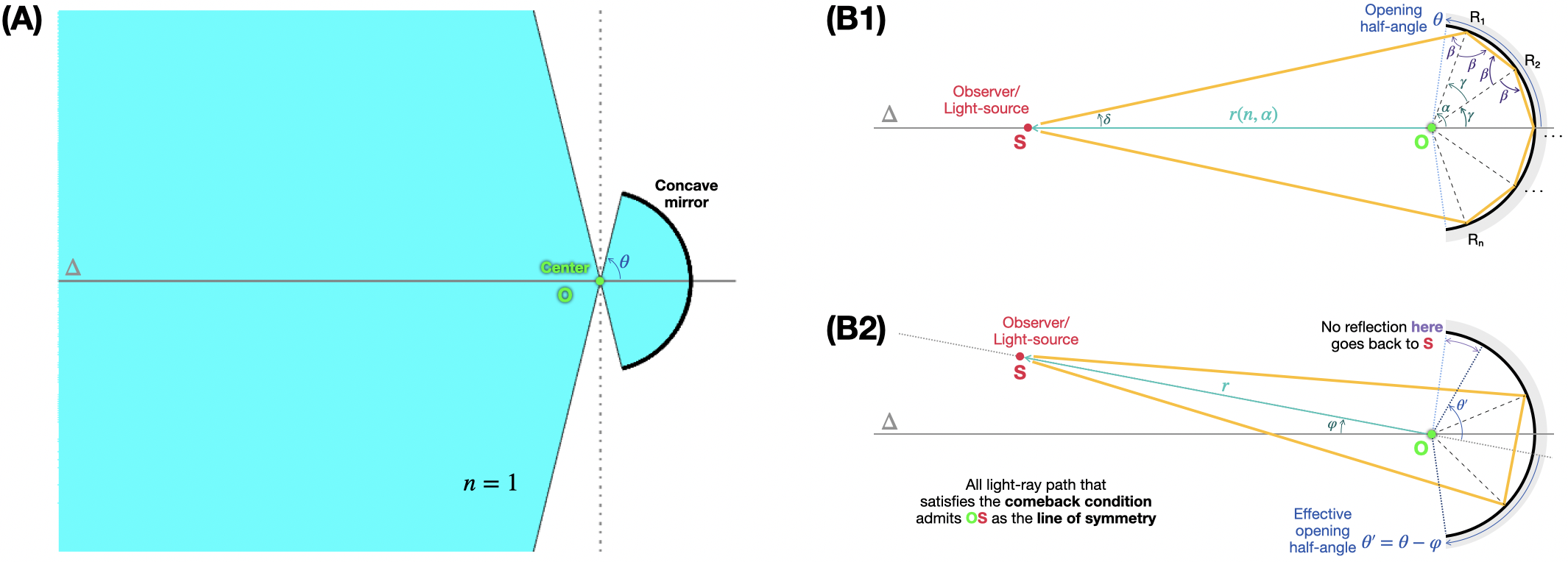}
\caption{{\color{black}\textbf{2D theory for the ray return condition.} In \textbf{(A)}, the cyan region shows all positions in space where a $n=1$ light-ray path exists. In \textbf{(B)}, we introduce the notation for analyzing higher-order $n>1$ light-ray paths: we represent the observer position $S$ in polar coordinates $(r,\varphi)$ with the origin at the center $O$ of the mirror’s circular arc. We consider positions along the the reflective symmetry axis $\Delta$ ($\varphi=0$) in \textbf{(B1)}, and off-axis ($\varphi \neq 0$) in \textbf{(B2)}.}}
\label{fig03}
\end{figure*}

We denote the center of the mirror's circular arc as O, the reflective symmetry axis of the mirror (in two-dimensional space) as $\Delta$, the half-opening angle of the mirror as $\theta$, and the point representing the observer as S -- see Fig. \ref{fig03}. For simplicity, in this work we only focus on $\theta \leq \pi/2$. We adopt polar coordinates $(r,\varphi)$ for the observer S, with the origin located at the center O of the mirror's circular arc; $r$ is the distance between points O and S, and 
$\varphi$ is the angle between the axis $\Delta$ and the line OS.

The condition for the existence of a $n=1$ light-ray path is that the ray retraces its path, which requires normal incidence on the mirror. Equivalently, the extension of the line segment OS must intersect the mirror and approach it from the reflective side. We show all positions in space that satisfy this condition in Fig. \ref{fig03}A.

For higher-order $n>1$ light-ray paths, let us split our investigation into two situations: the observer S is on the axis $\Delta$, or the observer S is off the axis $\Delta$. In other words, we consider the cases in which $\varphi=0$ (see Section \ref{sec:on}) and $\varphi \neq 0$ (see Section \ref{sec:off}).

\subsubsection{On-Axis Observer \label{sec:on}}

If there exists a $n>1$-order light-ray path that starts from the observer S then returns to S after $n$ reflections, then we seek to determine an expression for the distance $r(n,\alpha)$ between O and S, where $\alpha$ is the arc half-angle between the first and last reflection points, i.e. R$_1$ and R$_n$, respectively. Here, R$_1$, R$_2$, ..., R$_n$  represent the reflection points of the light-ray path on the mirror (see Fig. \ref{fig03}B1). The arc angle $\gamma$ between successive reflection points, i.e. R$_j$ and R$_{j+1}$, can be calculated with $2\alpha/(n-1)$, since the total arc between R$_1$ and R$_{n}$ is partitioned into $n-1$ equal intervals. The incidence angle $\beta$ is the same at every reflection point along the path, which can be found by looking at the sum of all angles in the triangle R$_1$OR$_2$:
\begin{equation}
\begin{split}
    \pi = \widehat{\text{R$_2$R$_1$O}} + \widehat{\text{R$_1$OR$_2$}} + \widehat{\text{ OR$_2$R$_1$}} = \beta + \gamma + \beta \ \
    \\
    \Longrightarrow \ \ \beta = \frac{\pi}{2} - \frac{\alpha}{n-1} \ .
\label{get_beta}
\end{split}
\end{equation}
We can then determine the angle $\delta$ between the in-coming light-ray and the axis $\Delta$ by looking at the sum of all angles in the triangle R$_1$OS:
\begin{equation}
\begin{split}
    \pi = \widehat{\text{SR$_1$O}} + \widehat{\text{R$_1$OS}} + \widehat{\text{ OSR$_1$}} = \beta + (\pi-\alpha) + \delta \ \
    \\
    \Longrightarrow \ \ \delta = -\frac{\pi}{2} + \frac{n \alpha}{n-1} \ .
\label{get_delta}
\end{split}
\end{equation}
We have found how to obtain all relevant angles in terms of $\alpha$. Now, let us look at the distance $h$ between the point R$_1$ and the axis $\Delta$, which can be calculated in two different ways:
\begin{itemize}
    \item From the center O with the viewing angle $\alpha$:
    \begin{equation}
        h = \mathcal{R} \tan \alpha \ .
    \label{h_way_1}
    \end{equation}
    \item From the observer S with the viewing angle $\delta$:
    \begin{equation}
        h = \left[ r(n,\alpha)\mathcal{R} \cos\alpha \right] \tan \delta \ .
    \label{h_way_2}
    \end{equation}
\end{itemize}
We can equate Eq. \eqref{h_way_1} and Eq. \eqref{h_way_2} then apply the expressions of angles found in Eq. \eqref{get_beta} and Eq. \eqref{get_delta} to arrive at the following formula:
\begin{equation}
    r(n,\alpha) = -\mathcal{R} \left[ \sin\alpha \tan \left( \frac{n\alpha}{n-1}\right) + \cos\alpha \right] \ .
\label{r_n_alpha_formula}
\end{equation}
Thus, increasing $\alpha$ leads to a decrease in $r(n,\alpha)$. Because the half-opening angle $\theta$ bounds $\alpha$ from above (i.e. $\alpha \leq \theta$), there is a minimal distance (from O and along $\Delta$) where
the $n>1$-order self-images first appear, which is given by $r(n,\theta)$. For any $\theta \leq \pi/2$, the required distance to start seeing more that one self-image is $r(2,\theta) \geq 0$.

If the maximum visible self-image order at a given point is $n>1$, then, since geometry does not forbid the visibility of lower order self-images, all lower-order self-images are also visible. Given that all higher-order self-images come in pairs (as mentioned before and illustrated in Fig. \ref{fig03}B2), the total number of visible self-images is $N=2n-1$.

From Eq. \eqref{r_n_alpha_formula}, we notice that the distance $r(n,\theta)$ reaches infinity when
$$\frac{n\theta}{n-1} \rightarrow \frac{\pi}{2} \ , $$
Therefore, the maximum order $n_{\max}$ of self-images can be seen as a function of the half-opening angle $\theta$:
\begin{equation}
    n_{\max}(\theta) = \left\lfloor \frac{\pi}{\pi - 2\theta} \right\rfloor \ ,
\label{n_max}
\end{equation}
where $\lfloor \frac{\pi}{\pi - 2\theta} \rfloor$ is the floor-function. In other words, for a concave mirror with a half-opening angle $\theta$, the axis $\Delta$ can be divided into intervals on which the visible self-image count takes odd values 1, 3, ..., $N_{\max}=2n_{\max}(\theta)-1$, i.e.
\begin{equation}
\begin{split}
    r\Big|_{\varphi=0} = 0 \xleftrightarrow{ \ N=1 \ } r(2,\theta) \xleftrightarrow{ \ N=3 \ } r(3,\theta) \xleftrightarrow{ \ N=5 \ }
    \\
    ... \xleftrightarrow{ \ N=2(n-1)-1 \ }  r(n,\theta) \xleftrightarrow{ \ N=2n-1 \ } ... 
    \\
    \xleftrightarrow{ \ N=N_{\max} \ } r\left( n_{\max}(\theta),\theta \right) \ .
\end{split}
\label{boundaries_of_line}
\end{equation}
The set $\{r(n,\theta)\}^{n_{\max}(\theta)}_{n=2}$ is the collection of points that partition the axis $\Delta$ into segments of different visible self-image counts $N$. The position $r(n>1,\theta)$ is the boundary point between $N=2(n-1)-1$ segment and $N=2n-1$ segment.

\subsubsection{Off-Axis Observer \label{sec:off}}

Not only do all higher-order self-images appear in pairs, owing to the reversibility of light-ray paths, but for reflections on a circular concave mirror, every light-ray path that satisfies the comeback condition is also symmetric about the line connecting the mirror’s center O and the observer S. Consequently, if the observer S is off-axis -- along a line with polar angle $\varphi \neq 0$ -- the number of self-images remains the same as in the on-axis case, except that the effective opening angle is modified to $\theta'=\theta-\varphi$ (see Fig. \ref{fig03}B2). Therefore, we can generalize the axis-partition in Eq. \eqref{boundaries_of_line} as the two-dimensional space partition, in which the boundaries between regions of different visible self-image counts $N$ -- say, region $N=2(n-1)-1$ and region $N=2n-1$, with $n>1$ -- are described by the curves:
\begin{equation}
\begin{split}
    &r(\varphi) = r(n,\theta') 
    \\
    & = - \mathcal{R} \left\{ \sin(\theta-\varphi) \tan \left[ \frac{n(\theta-\varphi)}{n-1}\right] + \cos(\theta-\varphi) \right\} \ ,
\label{r_n_theta_formula}
\end{split}
\end{equation}
where we have used Eq. \eqref{r_n_alpha_formula}. 

\begin{figure*}[htbp]
\centering
\includegraphics[width=\textwidth]{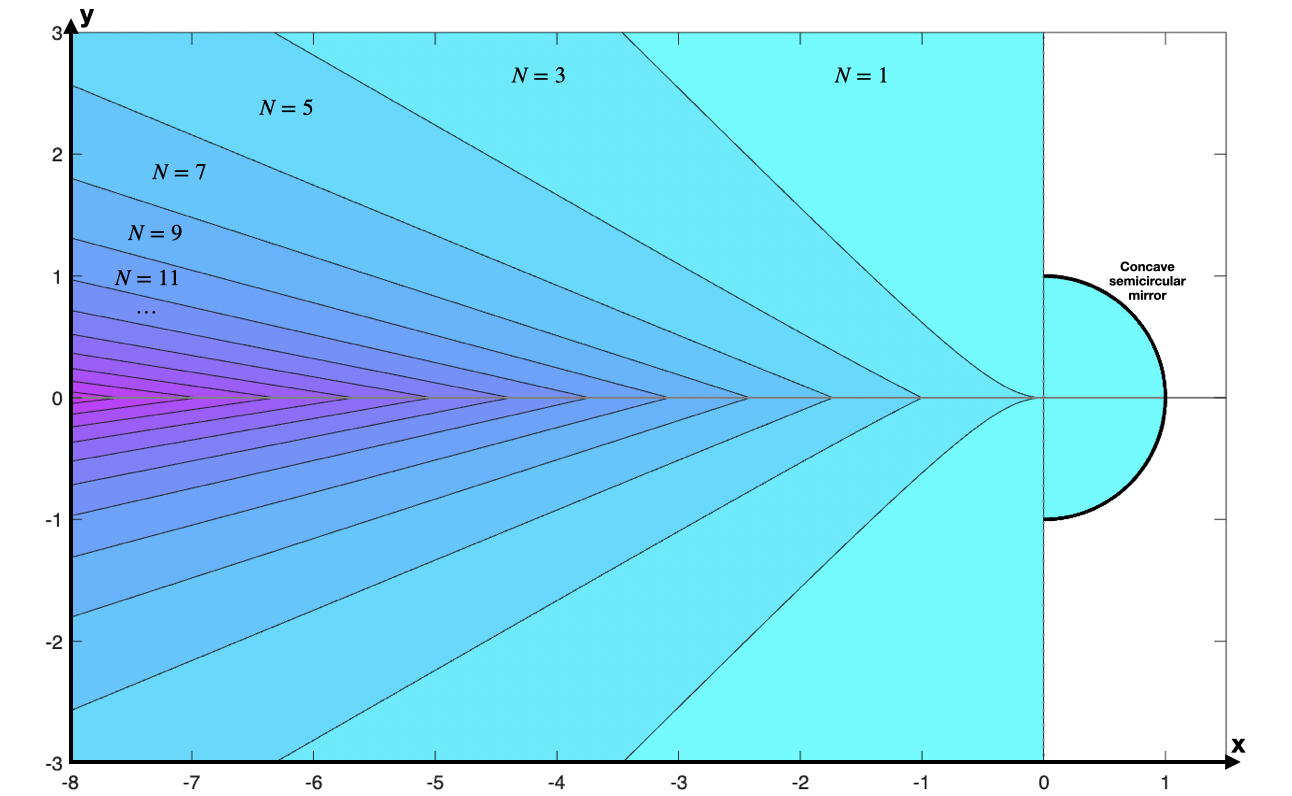}
\caption{\textbf{Spatial partition by self-image multiplicity for a semicircular concave mirror.} The half-opening angle of this mirror is $\theta = \pi/2$. Regions of the two-dimensional space are colored according to the number of self-images $N=1,3,5,...$ visible to the observer.}
\label{fig05}
\end{figure*}

Let us demonstrate the application of Eq. \eqref{r_n_theta_formula} with a concrete example. Consider a concave semicircular mirror, in which the half-opening angle is $\theta=\pi/2$. For convenience, we normalize the radius of the concave mirror to one unit, i.e. $\mathcal{R}=1$. Using Eq. \eqref{r_n_theta_formula}, we can determine all the spatial boundaries separating the regions that correspond to different numbers of self-images. As shown in Fig. \ref{fig05}, the two-dimensional space surrounding the semicircular concave mirror is thus divided into color-coded regions, each labeled by the number of self-images observable from that location. Furthermore, according to Eq. \eqref{n_max}, the space is partitioned into infinitely many such regions, since $n_{\max} \rightarrow \infty$.

\section{Three-Dimensional Experimental Results \label{sec:exp}}

\subsection{Methods}

To validate the theoretical findings in Section \ref{sec:theo}, we conducted an experiment with standard equipment from introductory laboratories. The concave cylindrical mirror we used\footnote{A cylinder concave stainless steel mirror, can be found at:\\ \url{https://www.eiscolabs.com/products/ph0529b}.} has radius $\mathcal{R} = 9.18 \pm 0.03$cm and half-opening angle $1.23 \pm 0.01$rad (see Fig. \ref{fig01}A), with measurement procedures detailed in Appendix \ref{app:measurement}.

In our experiment, the observer is a webcam\footnote{Logitech Brio 4K webcam, can be found at:\\ \url{https://www.logitech.com/en-us/shop/p/brio-4k.960-001390}.} and the light-source is a UV-glow sticker\footnote{Neasyth 3D adhesive phosphorescent dots, green}. By placing the sticker near the webcam lens (see Fig. \ref{fig01}B), we can approximate the sticker’s reflections as the observer’s self-images. We charge the sticker with a UV light source\footnote{Indmird UV black lights, 100 W (IP65)} and then acquire images in darkness so that the only illumination is the sticker’s phosphorescence. We print a 24''$\times$36'' poster at 300dpi showing the predicted spatial partitioning from Eq. \eqref{r_n_theta_formula}, place the mirror and webcam atop the poster, and record images\footnote{Taken by Zoom Version: 6.5.9 (61929).} at marked locations to sample regions of differing self-image multiplicity. The poster includes a Cartesian coordinate system $(x,y)$ in which the mirror’s center should be placed at the origin $(0,0)$, and the $\Delta$-axis is chosen to coincide with the $x$-axis (see Fig. \ref{fig01}C). When acquiring images, the webcam remains facing forward without turning toward the mirror, i.e. its optical axis always stays aligned to the $x$-axis.

There are a few practical notes for this experiment. When a self-image is generated by a ray reflecting very near either vertical edges of the mirror, spurious multiplicities can arise due to small departures from perfect circularity. These artifacts are often easy to identify and exclude, as they cluster unusually close together and appear visibly deformed relative to genuine images. A medium-sized UV-glow sticker\footnote{About 0.24'' in diameter.} can blur these clusters and thereby obscuring such defects, while still being small enough to tell non-defective self-images apart. Reflected light from the poster is also present; it can be easily identified and should not be counted as additional self-images.

\subsection{Results}

\begin{figure*}[htbp]
\centering
\includegraphics[width=\textwidth]{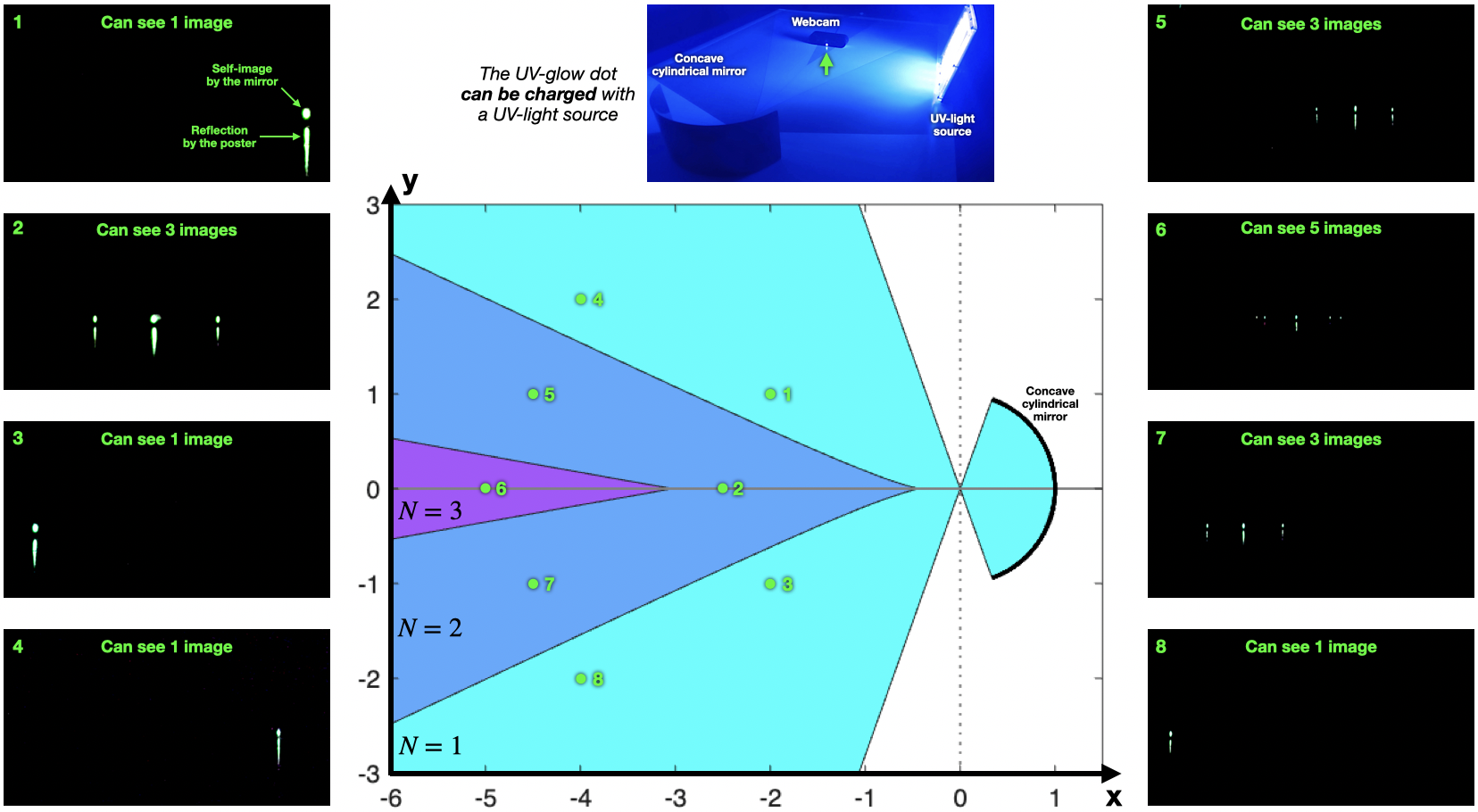}
\caption{\textbf{Imaging experiment with our concave cylindrical mirror.} We position the webcam at eight locations (marked in green, i.e. 1, 2, 3, ..., 8) and keep its optical axis fixed along the x-direction, then capture images in total darkness. The UV-glow sticker can be charged with a UV light source before imaging if needed. We show the captured images at these locations on the sides of this figure; the observed self-image multiplicity matches our theoretical prediction.}
\label{fig04}
\end{figure*}

On the poster, we acquired images at the following eight $(x,y)$ coordinates:
$$ (x,y) = (-2.5,0), (-2,\pm 1), (-5,0), (-4.5,\pm 1), (-4,\pm 2) \ . $$
This set of locations provides a sparse coverage of the poster area. Figure \ref{fig04} presents the imaging results at those locations; the observed self-image multiplicities match the theoretical predictions across all tested locations. 

{\color{black}There are a few things to consider regarding the differences between the idealized setting and a realistic experiment. Although a real reflector has reflectance less than $1$, for a reasonably good mirror and a small number of reflections, the attenuation and associated blur are modest, so the high-order images (but not too high) remain observable. The main practical deviations arise from surface imperfections (e.g. small bumps or waviness), which perturb the local normals and can deform the image or even split it into multiple very close images (a cluster of images). However, if the light source is sufficiently large, these split images overlap and become indistinguishable, and thus still appear as a single feature.}

\section{What have We learned?}

Using only elementary ray-optics and symmetry arguments, we have quantified how a concave half-cylindrical mirror produces self-image multiplicity. We then have verified our theoretical findings with simple experiments, which are at the level of introductory lab. Our work shows that seemingly routine laboratory setups can conceal rich, quantitative insights and highlights the value of curiosity-driven investigation. {\color{black} Because \textit{high-frequency} sound propagation admits a geometric (ray tracing) description with specular reflection/refraction laws analogous to geometric optics, our results may also inform geometric acoustic simulations \cite{savioja2015overview,de2015modeling,vorlander2013computer} -- in particular, the analysis of \textit{specular reflections} (e.g. wall/object reflections) \cite{vorlander1989simulation,kuster2006modelling} and \textit{caustic focusing} (e.g. localized sound “hot spots”) \cite{porter1987gaussian,mo2017outdoor} as recorded by finite receiver apertures or microphone arrays.}

\section{Acknowledgement}

We would like to thank Adam Landsberg for many guidance and discussion during the early experimental stage of this project. We also would like to thank Tiziana Di Luccio, Gloria Lee, and Stephen Simon for matters related to introductory to physics lab.

\bibliography{main}
\bibliographystyle{apsrev4-2}

\appendix

\section{Cylindrical Concave Mirror Measurements \label{app:measurement}}

\begin{figure*}[htbp]
\centering
\includegraphics[width=\textwidth]{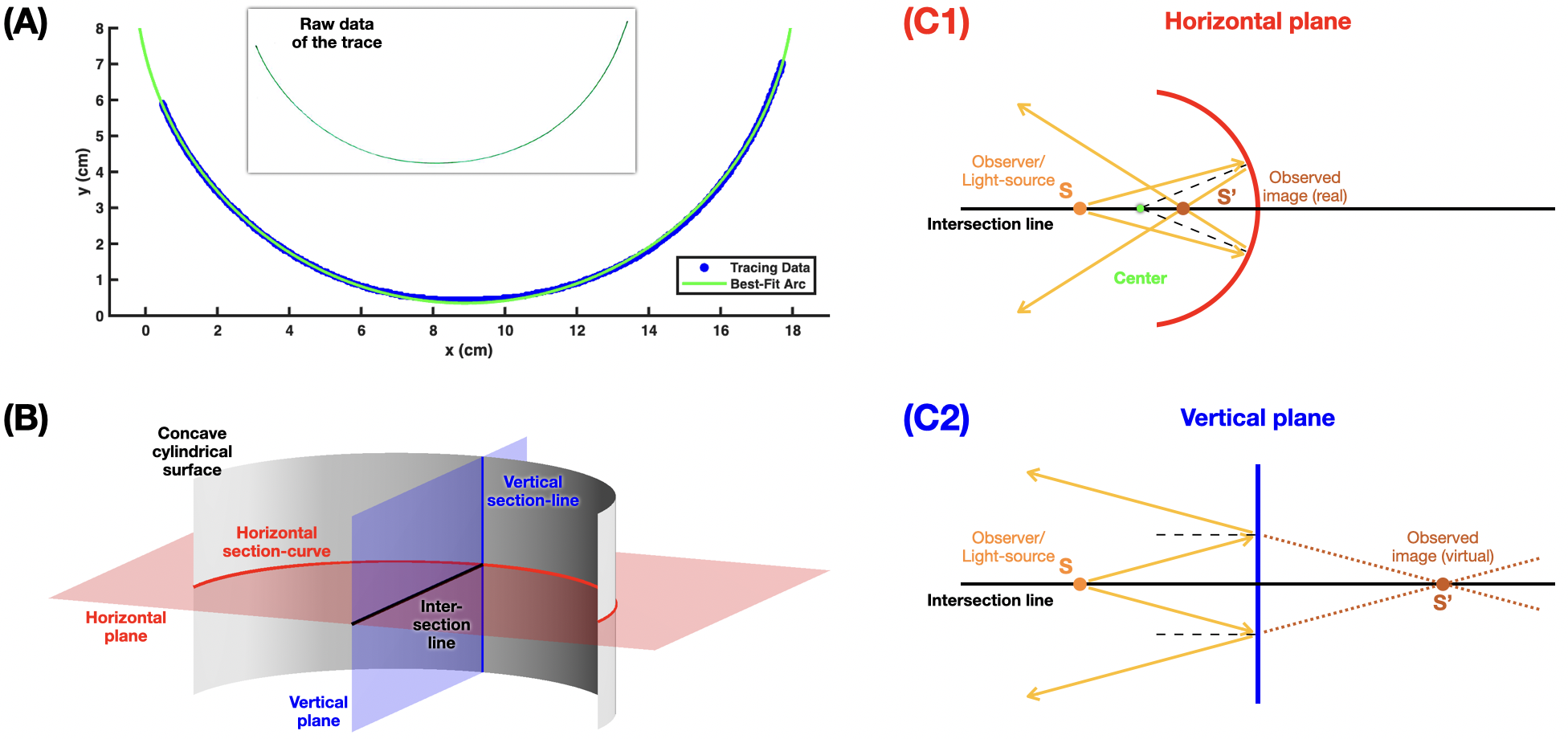}
\caption{\textbf{Geometry and measurements of our concave cylindrical mirror.} The cross-section of our mirror, normal to the cylinder axis, is a circular arc. \textbf{(A)} The horizontal section-curve of our mirror, represented in the Cartesian $(x,y)$-coordinate system, precisely follows a circular arc. The inset shows the raw edge-trace of the mirror, drawn with green ink. \textbf{(B)} The three-dimensional geometry of our mirror in which we specify a horizontal plane and a vertical plane for further investigation. \textbf{(C1)} The reflective behavior of light-rays confined to the horizontal plane. \textbf{(C2)} The reflective behavior of light-rays confined to the vertical plane.}
\label{figS01}
\end{figure*}

To perform our experiment, we first need to determine how closely our concave mirror approximates a cylindrical surface, then estimate its radius $\mathcal{R}$ and half-opening angle $\theta$. We place the cylindrical mirror on a white sheet of paper and trace its entire edge using a green-ink pen (see Fig. \ref{figS01}A inset); the distance between the two endpoints of the trace was measured to be $D=17.30 \pm 0.05$cm. After scanning the paper, we isolate the green channel and apply thresholding to identify the traced edge, converting it into a collection of $(x,y)$-coordinate points (see Fig. \ref{figS01}A). We then use the curve-fitting toolbox \textit{cftool} in \textit{MATLAB R2025a} \cite{MATLAB-R2025a} to curve-fit these points using the following function:
\begin{equation}
    y = y_0 - \left[\mathcal{R}^2 - (x-x_0)^2 \right]^{1/2} \ ,
\end{equation}
where $x_0$, $y_0$, and $\mathcal{R}$ are curve-fitting parameters. The fitted curve shows excellent agreement with the data, yielding an adjusted $R^2$-value\footnote{also known as the metric for \textit{goodness of the fit}.} of $0.9994$ (very close to $1$, which is the corresponding value for a perfect fit). If we take $D=17.30$cm exactly, the best-fit value of $\mathcal{R}$ is $9.18 \pm 0.01$cm (at $95\%$ confidence level). Combining this with the measurement uncertainty of $D$, we obtain the estimation for the cylindrical radius $\mathcal{R}=9.18 \pm 0.03$cm. For the half-opening angle $\theta$, we apply the equation:
\begin{equation}
    \theta = \arcsin\left( D/2\mathcal{R} \right) \ ,
\end{equation}
and get the estimation $\theta = 1.23 \pm 0.01$rad. 

\section{Astigmatism \label{app:astigmatism}}

To see an image through a mirror, a \textit{bundle} of nearby light-rays must leave the object, reflect off the mirror's surface (possibly multiple times -- the number indicates the image \textit{order}), and reach the eye. An image is perceived at the point where the reflected rays either physically intersect (forming a \textit{real} image) or would intersect if extended backward (forming a \textit{virtual} image); as long as enough of that ray bundle enters the pupil and the convergent point lies within the eye's focusing range -- that is, in front of the eye -- the eye's lens will project a sharp image onto the retina. However, in three dimensions, reflected rays do not always converge to a single sharp point. A common optical distortion known as \textit{astigmatism} can occur, causing the image to appear blurry or stretched along one direction \cite{hecht2012optics}. This behavior happens with any cylindrical concave mirror, where reflected rays converge differently in the \textit{horizontal} and \textit{vertical} planes, resulting in non-coinciding \textit{transverse} and \textit{sagittal} images.

Let us take a look at the surface geometry of our concave cylindrical mirror, and consider a horizontal plane and a vertical plane as illustrated in Fig. \ref{fig01}B. The intersections between these planes and the mirror are a horizontal section-curve (which is a circular arc of radius $\mathcal{R}$) and a vertical section-line (which is a straight line), respectively. In Fig. \ref{figS01}C1 and Fig. \ref{figS01}C2, we analyze the two planes and the reflective behavior for light from a source $S$ located in front of the mirror on the intersection line of these planes. The observed image (convergent point) by a light bundle centered on the intersection line may be real in the horizontal plane, but is always virtual in the vertical plane and,  more importantly, will always appear in front of the observer, thereby satisfying the \textit{visibility condition}.

\end{document}